\documentclass[aps, prd, preprintnumbers, floatfix, showpacs, showkeys, nofootinbib, 10pt]{revtex4-1}

\usepackage[dvips]{graphics}
\usepackage{amsmath,amsfonts,bm}
\usepackage{amssymb}
\usepackage{amsthm}
\usepackage{epsfig,bm}
\usepackage{feynmp}
\usepackage{graphicx,comment}
\usepackage{dcolumn,color}
\usepackage{mathtools}
\usepackage{fancyhdr}
\usepackage{bm}
\usepackage{cancel}
\usepackage{slashed}
\usepackage{verbatim}
\usepackage{listings}
\usepackage{graphicx,latexsym}
\usepackage{rotating}
\usepackage{color}
\usepackage{float}
\usepackage{enumerate}
\usepackage{array}
\usepackage{tabularx}
\usepackage{longtable}
\usepackage{booktabs}

\def\beq{\begin{equation}}
\def\eeq{\end{equation}}
\def\bea{\begin{eqnarray}}
\def\eea{\end{eqnarray}}

\begin{document}
\preprint{\today}


\title{Chain Formation in a 2-Dimensional System of Hard Spheres with a Short Range, Anisotropic Interaction}


\author{Nelia~Mann$^1$}
\affiliation{$^1$ Union College, Schenectady NY, 12308, USA}


\begin{abstract}

We analyze a generalization of the hard sphere dipole system in two dimensions in which the interaction range of the interaction can be varied.  We focus on the system in the limit the interaction becomes increasingly short-ranged, while the temperature becomes low.  By using a cluster expansion and taking advantage of low temperatures to perform saddle-point approximations, we argue that a well defined double limit exists in which the only structures which contribute to the free energy are chains.  We then argue that the dominance of chain structures is equivalent to the dominance of chain diagrams in a cluster expansion, but only if the expansion is performed around a hard sphere system (rather than the standard ideal gas).  We show that this leads to non-standard factorization rules for diagrams, and use this to construct a closed-form expression for the free energy at low densities.  We then compare this construction to several models previously developed for the hard sphere dipole system in the regime where chain structures dominate, and argue that the comparison provides evidence in favor of one model over the others.  We also use this construction to incorporate some finite density effects though the hard sphere radial distribution function, and analyze the impact of these effects on chain length and the equation of state.

\end{abstract}



\maketitle


\section{Introduction}

The study of particles which interact anisotropically has a rich history.  Models developed to explore these systems have included hard-sphere dipoles, soft-sphere dipoles, Stockmayer fluids, and theories of particles that are spherocylindrical, disk-like, and dumbbell in shape.  This theoretical work has been supplemented by inreasingly sophisticated computational approaches, as wekk as experimental work on colloidal suspensions of magnetic particles, which though necessarily more complex than the simple theoretical models, nonetheless showed many of the same properties and phenomena.

Even the simplest theoretical model for describing such systems, that of the Dipolar Hard Sphere (DHS), gives rise to a variety of phases and phenomena and has provided several surprises during its study.  At high densities, crystal solid phases exist as well as both isotropic and ferromagnetic fluids \cite{WeisLevesqueZarra, WeisLevesque1993, GrohDietrich}.  At low densities, early theoretical work using thermodynamic perturbation theory, the mean spherical model,  as well as integral equations suggested many similarities between the DHS system and the more familiar hard-sphere Van der Waals system (see for example \cite{DeGennesPincus, Wertheim71, RushbrookeStellHoye}).  Indeed, these approaches generally involved a thermodynamic average over the anisotropic $\sim \frac{1}{r^3}$ interaction, which results in the more familiar, isotropic $\sim \frac{1}{r^6}$ attraction.  On the basis of these efforts it was believed for a long time that the DHS system would undergo a gas-liquid phase transition extremely similar to the one that appears in the Van der Waals system.  

Though early Monte Carlo simulations at high temperatures seemed in good agreement with this theoretical work \cite{VerletWeis, Patey, NgValleauTorriePatey}, increasingly sophisticated and systematic efforts failed to find any phase transition under the conditions predicted in the DHS system \cite{Caillol}, and related systems (such as those with an additional isotropic Van der Waals interaction or those with non-spherical hard particles) were shown to lose such a phase transition as they approached the DHS system \cite{vanLeeuwenSmit, McGrotherJackson}.  Instead, it was discovered that as the temperature decreased, chains and rings formed, and that these objects interacted with each other only very weakly (as evidenced by only slight changes in the internal energy of the system with large changes in its density) \cite{WeisLevesque, LevesqueWeis}.  Experimental evidence analyzing magnetic colloids also identified the prevalence of chain structures \cite{WenKunPalZhengTu, KlokkenburgDullensKegelErnePhilipse}.  As a result, theoretical models based on chains were developed and used to study the energy density, distribution of chain lengths, and equation of state for the hard sphere system, as well as to analyze the possibility of a gas-liquid phase transition.

In \cite{Sear}, Sear analyzed a system of chains using a cluster expansion, taking into account only the contributions from chain diagrams.  In doing so, we was able to compute the energy density of the system and show that there was a large range of densities over which it would vary only weakly.  In \cite{OsipovTeixeiraTelodaGama, TavaresTelodaGamaOsipov}, Osipov, Teixeira and Telo da Gama created a different model for the hard sphere dipoles as ``living polymers'', in which the free energy of the system was written as a sum of terms coming from non-interacting chains of different lengths and was then minimized to determine the distribution of chain lengths.  Later, this model was modified in \cite{TavaresWeisTelodaGama1999} by treating individual monomers as indistinguishable; this treatment is not standard in polymer models, but improved agreement between the model and numerical results.  Further work based on the chain and polymer models included weak interactions between chains and analyzed how these models would suppress a gas-liquid phase transition \cite{Roij, Levin}.

It then came as a fresh surprise when evidence emerged in Monte Carlo simulations that a phase transition occurs after all \cite{CampShelleyPatey}, though at much lower temperatures and densities than those predicted by the early perturbative work.  In \cite{TlustySafran}, it was argued that this phase transition is associated with the formation of a network of chain segments connected via 3-way vertices, and that it could be explained by treating both the finite ends of chains and the vertices as defects in infinitely long chains; both defects were entropically favorable, with the vertices being energetically preferred over the finite ends, so that lowering the temperature drove network formation.  Later, a calculation using thermodynamic perturbation theory for associative fluids was also able to achieve good qualitative agreement with the observed phase behavior of the system at low temperatures and densities \cite{Kalyuzhnyi}.

Here, we are interested in exploring the role that interaction range plays in this story.  We will consider a 2-dimensional system of particles of diameter $\sigma$, which interact via both a hard-sphere repulsion and an additional generalized anisotropic pair-wise potential of the form
\beq
\label{eqn:pairwise}
u_{12} = \frac{u_0 \, \sigma^{p}}{2r_{12}^{p}} \, \Big[\hat{m}_1\cdot\hat{m}_2 - 3(\hat{m}_1\cdot\hat{r}_{12})(\hat{m}_2\cdot\hat{r}_{12})\Big] \, ,
\eeq
where $p$ is an arbitrary constant, $\vec{r}_{12}$ is the displacement vector between the two particles, and $\hat{m}_i$ gives the orientation of the dipole for each particle.  The choice to work in 2-dimensions here is made for calculational convenience; previous work has found that a ``pseudo-2-dimensional'' DHS system, with particles and dipoles confined to a plane but with $p = 3$  (a model for a ferromagnetic monolayer) behaves similarly to the 3-dimensional DHS system \cite{Weis, LombaLadoWeis, WeisTavaresTelodeGama, TavaresWeisTelodaGama2002, TavaresWeisTelodaGama, DuncanCamp, GeigerKlapp}.

Although systems with $p \ne 3$ would arise naturally in a self-consistent electromagnetic theory in $D$ dimensions, here we will take a more pragmatic view and think of $p$ as simply a constant that controls how short- or long-ranged the interaction between particles is.  We are particularly interested in a limit where $p$ becomes large.  At finite temperature, as $p \rightarrow \infty$ the anisotropic interaction simply disappears.  However, if we consider a double limit where the temperature becomes small and $p$ becomes large, we enter a regime which might be thought of as a ``nearest-neighbor'' approximation to the original theory.  Furthermore, we will show that while making $p$ large decreases the internal entropy of all types of particle arrangements, it affects vertices more than it does chains, so that as $p \rightarrow \infty$ and $\tau \rightarrow 0$, we can create a theory in which only chains of particles form, and in which the chains themselves are treated using a nearest-neightbor approximation.  The chain/polymer models analyzed in \cite{Sear, OsipovTeixeiraTelodaGama, TavaresTelodaGamaOsipov, TavaresWeisTelodaGama1999} typically neglect all but nearest-neighbor interactions when modeling the interiors of these chains, and either entirely neglect interactions between chains or treat them as perturbative corrections, so our construction can be thought of as creating a formal framework in which these previous approximations hold.  As a result, calculations within this model can be used to gain insight into the approximations made in these earlier DHS models.  Previous work in \cite{OsipovTeixeiraTelodaGama2} has also explored the implications of short- and long-ranged interactions with anisotropic interactions.

In section \ref{generalities}, we will analyze this system using a cluster expansion of the free energy.  Normally cluster expansions use an ideal gas as the reference system and work at low densities, but we will instead be perturbing around a hard sphere system.  At low temperatures, each term in the density expansion is dominated by arrangements of particles that represent local energy minima, and we can use a saddle point approximation to determine the contributions from these arrangements.  For large $p$, we must also take into account how the internal entropy of these arrangements scales with $p$.  We will show that a double limit should exist in which the only contributions which survive are those associated with chain structures, and that the contributions from these arrangements arise entirely from chain diagrams.

In section \ref{chains}, we will compute the contributions from these structures and analyze the factorization properties of these diagrams.  By comparing the results to the standard factorization rules that apply when we expand around an ideal gas, we conclude that it does not make sense to restrict the expansion to chain diagrams unless this expansion is performed using the hard sphere system as a reference.  At low densities, we can use the factorization properties to find a closed-form expression for the free energy of the system.  

In section \ref{comparisons}, we discuss how our construction gives insight into three chain/polymer models for the DHS system \cite{Sear, OsipovTeixeiraTelodaGama, TavaresTelodaGamaOsipov, TavaresWeisTelodaGama1999}.  We argue that our results are equivalent to those reached in the polymer model developed in \cite{OsipovTeixeiraTelodaGama, TavaresTelodaGamaOsipov}, which provides evidence in favor of it.  In section \ref{finitedensity}, we argue that with additional assumptions regarding the factorization properties of hard sphere distribution functions, we can extend our results to include some finite density effects, and discuss their implications.  In section \ref{conclusions}, we discuss our results and possible future research directions.  Appendix \ref{combinatorics} includes the combinatorial argument used to connect factorization rules for chain diagrams to the calculation of the free energy.

\section{Cluster Expansion Diagrams and the Saddle Point Approximation}
\label{generalities}

We will begin by briefly reviewing the set-up of a cluster expansion, following roughly \cite{Goodstein} but perturbing around hard spheres.  Let us suppose that we wish to relate the partition function of our system $Z$ to the partition function of the hard sphere system $Z_{\mathrm{HS}}$, so we write
\beq
Z(\eta, \tau) = Z_{\mathrm{HS}}(\eta) \langle e^{-U/\tau} \rangle_{\mathrm{HS}} 
\eeq
where if we have $N$ particles confined to area $A$, then $\eta = \frac{\pi \sigma^2 N}{4A}$ is the packing fraction, $\tau = k_BT$, $U$ is the total energy associated with the anisotropic interaction, and $\langle \cdots \rangle_{\mathrm{HS}}$ represents a thermal average over the hard sphere system.\footnote{Cluster expansions are usually applied to the grand partition function, where factorization properties of the diagrams leads to significant simplifications.  However, as will be discussed in the next section, we will not be able to rely on these properties, so it is more convenient to use the partition function.}  We can then use Mayer functions $f_{ij} = e^{-u_{ij}/\tau} - 1$ to write
\beq
\langle e^{-U/\tau} \rangle_{\mathrm{HS}}  = \left\langle \prod_{i < j} (1 + f_{ij})\right\rangle_{\mathrm{HS}} =1 + = 1 + \frac{N!}{2(N-2)!}\langle f_{12}\rangle_{\mathrm{HS}} + \frac{N!}{3!(N-3)!}\langle f_{12}f_{23}f_{13} + 3f_{12}f_{23}\rangle_{\mathrm{HS}}  + \cdots 
\eeq
where we are grouping terms by the number of particles involved in the calculation.  If we then define $\left(\frac{8\eta}{N}\right)^{m-1}D_m$ as the combination, including symmetry factors, of all diagrams associated with $m$ particles (so that the object $D_m$ is a dimensionless integral), we arrive at
\beq
\label{eqn:Dmexpansion}
\langle e^{-U/\tau} \rangle_{\mathrm{HS}} = 1 + \sum_{m = 2}^{N} \frac{N!}{m!(N-m)!}\left(\frac{8\eta}{N}\right)^{m-1} \, D_m \, ,
\eeq
where
\beq
\label{eqn:Dm}
D_m = \left[\prod_{i = 2}^{m} \int \frac{d^2\vec{x}_i}{2\pi} \right]\left[\prod_{i = 1}^{m} \int\frac{d\theta_i}{2\pi}\right] g(\{\sigma \vec{x}_i\}; \eta) \ \Big( e^{-\tilde{U}_{m}/\tilde{\tau}} + \cdots\Big) \, .
\eeq
Here,  $\vec{x}_i = \vec{r}_i/\sigma$ are the dimensionless positions, and the functions $g(\{\vec{r}_i\}; \eta)$ are hard sphere distribution functions.  The energy $\tilde{U}_m = \frac{U_m}{u_0}$ is the total energy of the arrangement of $m$ particles, expressed in the natural units of the system.  In the limit that $\tilde{\tau} = \frac{\tau}{u_0}$ becomes small, this integral is dominated by physical arrangements of particles that represent local minima of $\tilde{U}_{m}$, and the terms we are suppressing only matter in that they impose the condition that each particle must have a non-zero interaction energy with at least one other particle in every minimum we consider, as was gauranteed by the original construction in terms of Mayer functions.  

Suppose we use an index $k$ to denote energy minimizing arrangements of $\tilde{U}_{m}$ which are distinct under relabeling of particles, and define $D_{(m;k)}$ as the contribution to $D_m$ that comes from expanding the integrand in equation \ref{eqn:Dm} near this minimum.  We might write $D_m \approx \sum_{k} D_{(m;k)}$, but in order for the approximation we are making to be self-consistent, we should only keep the dominant term(s) in this sum.  For fixed $p$ and $\tilde{\tau} \rightarrow 0$ this contribution comes from the global minimum of $\tilde{U}_m$, but when we have $p \rightarrow \infty$ and $\tilde{\tau} \rightarrow 0$ other terms can dominate.

The integral we must perform to compute one of the $D_{(m;k)}$ involves $m-1$ integrals over $2$-dimensional positions, and $m$ integrals over orientation angles, with one spatial integral (corresponding to rotations of the entire arrangement) being trivial.  Suppose for a particular local minimum, there are $m_c$ points of contact between particles.  It is then convenient to use a set of variables $\varepsilon_{\alpha}$, for $\alpha = 1, \dots, m_c$, with $1 + \varepsilon_{\alpha}$ being the distance between a pair of particles which are in contact at that minimum.  If the minimum is degenerate with moduli space $M$, we can parametrize it using one or more moduli, written as a $m_m$ component vector $\vec{\zeta}$.  For the remaining variables of integration, we can use any set of $\psi_{a}$, with $a = 1, \dots, m_f$,  defined so that $\psi_a = 0$ at the minimum.  In this case, we can write
\beq
\label{eqn:expansion}
D_{(m;k)} = \frac{C \, e^{-\tilde{U}_{0}/\tilde{\tau}}}{(2\pi)^{2(m-1)}} \, \int_{M} d\vec{\zeta} \, \Big|J_0(\vec{\zeta})\Big| \, g_0(\vec{\zeta}; \eta) \left[\prod_{\alpha = 1}^{m_c}\int_{0}^{\infty} d\varepsilon_{\alpha}\right]\left[\prod_{a = 1}^{m_f}\int_{-\infty}^{\infty} d\psi_{a} \right] \, \mathrm{Exp}\left[-\frac{1}{\tilde{\tau}}\left(K^{\alpha}(\vec{\zeta})\varepsilon_{\alpha} + \frac{1}{2}H^{ab}(\vec{\zeta})\psi_{a}\psi_{b} + \cdots\right)\right] \, .
\eeq
Here, $C$ is the number of minima equivalent under relabeling of particles.  $\tilde{U}_{0}$ represents the value of $\tilde{U}_{m}$, $|J_0(\vec{\zeta})|$ represents the magnitude of the Jacobian determinant arising from the variable transformation $\{\vec{x}_i, \theta_{i}\} \rightarrow \{\vec{\zeta}, \varepsilon_{\alpha}, \psi_a\}$ and $g_{0}(\vec{\zeta}; \eta)$ represents the hard-sphere distribution function, all evaluated at the minimum.  The objects $K^{\alpha}(\vec{\zeta})$ and $H^{ab}(\vec{\zeta})$ are defined in terms of the gradient and Hessian of $\tilde{U}_{m}$, again evaluated at the minimum:
\beq
K^{\alpha}(\vec{\zeta}) = -\left.\frac{\partial \tilde{U}_{m}}{\partial\varepsilon_{\alpha}}\right|_{0}, \hspace{1in} H^{ab}(\vec{\zeta}) = -\left.\frac{\partial^2 \tilde{U}_{m}}{\partial\psi_a\partial\psi_b}\right|_{0} \, .
\eeq
The fact that we are at a local minimum means we should have $K^{\alpha} > 0$, and $\det{H} \ge 0$, and if $\det{H} = 0$, we should extend the expansion in $\psi_{\alpha}$ to higher order in order to capture the leading low-temperature behavior.  Equation \ref{eqn:expansion} is of course the Boltzmann factor $e^{-\tilde{U}_{0}/\tilde{\tau}}$ multiplied by an expression determined by the internal entropy of the arrangement.  It scales as a power of $\tilde{\tau}$ that is determined by the structure of the arrangement, but in addition may depend on $p$ through $K$ and $\det H$.

As $p$ becomes large, we have $\tilde{u}_{ij} \rightarrow 0$ for any pair of particles not physically touching, so that the object $\tilde{U}_m$ includes only nearest neighbor interactions.  In this case, when expanding near a particular minimum we can write
\beq
\tilde{U}_m = -\sum_{\alpha = 1}^{m_c} \frac{\Phi_{\alpha}(\vec{\zeta}, \psi_{a})}{(1 + \varepsilon_{\alpha})^{p}} \, ,
\eeq
where the functions $\Phi_{\alpha}$ do not depend on either $p$ or the variables $\varepsilon_{\alpha}$.  In this case, $H^{ab}$ should be independent of $p$, and $K^{\alpha}$ should scale linearly with $p$ for each $\alpha$.  Thus, our expression for $D_{(m;k)}$ becomes 
\beq
\label{eqn:expansion2}
D_{(m;k)}  \ \stackrel{\longrightarrow}{\scalebox{0.5}{$p \rightarrow \infty$}} \  \frac{C \, e^{-\tilde{U}_{0}/\tilde{\tau}}}{(2\pi)^{2(m-1)}} \, \left(\frac{\tau}{p}\right)^{m_c} \, \int_{M} d\vec{\zeta} \, \frac{\Big|J_0(\vec{\zeta})\Big| \, g_0(\vec{\zeta}; \eta)}{\Phi_0(\vec{\zeta})} \left[\prod_{a = 1}^{m_f}\int_{-\infty}^{\infty} d\psi_{a} \right] \, \mathrm{Exp}\left[-\frac{1}{\tilde{\tau}}\left(\frac{1}{2}H^{ab}(\vec{\zeta})\psi_{a}\psi_{b} + \cdots\right)\right] \, ,
\eeq
where $\Phi_0(\vec{\zeta}) = \prod_{\alpha = 1}^{m_c} \Phi_{\alpha}(\vec{\zeta}, 0)$.  Also important is that in this limit arrangements of particles consisting of multiple, separated clusters can be valid minima (with large moduli spaces representing the relative positions and orientations of the separate clusters), and in fact can contribute significantly to the result.

Now consider what happens if we simultaneously make $p$ large and $\tilde{\tau}$ small.  If we think of an arrangement of particles as a set of nearest-neighbor interactions, each $\tilde{u}_{ij}$ contributing to $\tilde{U}_m$ satisfies $\tilde{u}_{ij} \ge -1$, so that $\tilde{U}_m \ge -m_c$.  Since only head-to-tail arrangement optimize the energy, we have $\tilde{U}_m = -m_c$ only when all nearest-neighbor interactions are head-to-tail (so the arrangement consists entirely of chains.)  On the other hand, each nearest-neighbor interaction generates a factor of $p^{-1}$ in $D_{(m;k)}$, so the expression $D_{(m;k)}$ will scale as $\sim \left(\frac{1}{p}e^{1/\tilde{\tau}}\right)^{m_c}$ for chain arrangements, and will be strongly suppressed for all others.  As a result, it is possible to define a formal double limit where the only arrangements which contribute are the chains.

We stress here that it is the directional nature of the interaction that allows for the creation of this double limit and the physical dominance of chain structures within it.  It would be possible to further generalize the interaction potential to the form
\beq
u_{12} = \frac{u_0 \sigma^{p}}{2r_{12}^{p}}\left[\hat{m}_1\cdot\hat{m}_2 - \kappa(\hat{m}_1\cdot\hat{r}_{12})(\hat{m}_2\cdot\hat{r}_{12})\right] \, ,
\eeq
and for any $\kappa > 2$ the head-to-tail arrangement of particles would be energetically favorable, and the same essential limiting behavior would be recovered.  Increasing the value of $\kappa$ then creates a system where chain dominance is reached more rapidly in the double limit.

This analysis somewhat obscures the connection between these terms and the original diagrammatic expansion.  In the large $p$ limit, the vanishing of $\tilde{u}_{ij}$ for any pair of particles not in contact also implies the vanishing of the associated Mayer function $f_{ij}$, so that the contribution to $D_m$ from an arrangement of chains can only arise in the equivalent diagram.  (For example, the contribution to $D_4$ from a chain of four particles comes from the diagram $\langle f_{12}f_{23}f_{34}\rangle_{\mathrm{HS}}$, while the contribution from two separate chains of two particles comes from the diagram $\langle f_{12}f_{34}\rangle_{\mathrm{HS}}$.)  This allows us to ignore any diagrams not composed entirely of chains, and write 
\beq
\langle e^{-U/\tau}\rangle_{\mathrm{HS}} \approx 1 + \frac{N!}{2(N-2)!}\langle f_{12}\rangle_{\mathrm{HS}} + \frac{3 \, N!}{3!(N-3)!}\langle f_{12}f_{23}\rangle_{\mathrm{HS}}  + \frac{N!}{4!(N-4)!}\Big(12\langle f_{12}f_{23}f_{34}\rangle_{\mathrm{HS}} + 3\langle f_{12}f_{34}\rangle_{\mathrm{HS}}\Big) + \cdots \, ,
\eeq
where each diagram is analyzed using the same saddle point approximation method.

\section{Chain Diagrams, Factorization, and the Free Energy}
\label{chains}

Next we want to analyze the chain diagrams, as well as the diagrams composed of multiple separated chains.  Here we will use the short-hand notation $\langle \{m\}\rangle_{\mathrm{HS}}$ for a chain diagram involving $m$ particles, and $\langle \{m_1, m_2, \dots, m_s\}\rangle_{\mathrm{HS}}$ for a diagram with $s$ separate chains of lengths $\{m_1, \cdots, m_s\}$.  

\subsubsection{The 2-Particle Chain}

\begin{figure}
\begin{center}
\resizebox{1.5in}{!}{\includegraphics{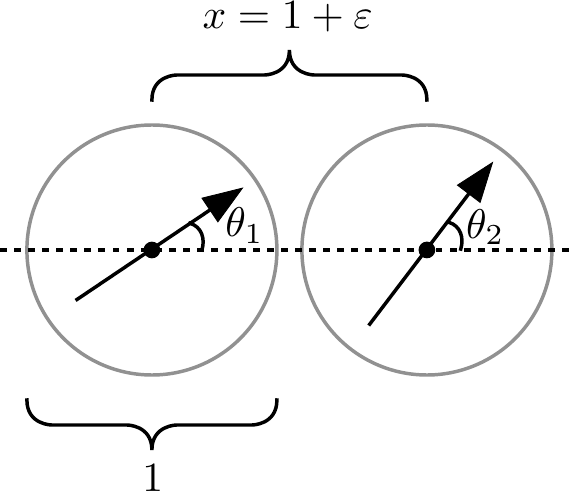}}
\caption{\label{twochain} A diagram labeling the variables used to describe an arrangement of two particles in the calculation of $\langle \{2\}\rangle_{\mathrm{HS}}$.}
\end{center}
\end{figure}

The calculation of $\langle \{2\}\rangle_{\mathrm{HS}} = \langle f_{12} \rangle_{\mathrm{HS}}$ differs from the well-known result for the second virial coefficient for DHS at low temperatures only in rather trivial ways, but we repeat it here for clarity.  The integral over parameter space is best expressed in terms of the variables $\{x, \theta_1, \theta_2\}$, shown in figure \ref{twochain}.  This gives us the expression
\beq
\langle \{2\} \rangle_{\mathrm{HS}} = \left(\frac{8\eta}{N}\right) \, \int_{1}^{\infty} g(\sigma x; \eta) \, dx \int_{0}^{2\pi}\int_{0}^{2\pi} \left(\mathrm{Exp}\left[\frac{2\cos\theta_1\cos\theta_2 - \sin\theta_1\sin\theta_2}{2\tilde{\tau}x^{p}}\right] - 1\right) \, \frac{d\theta_1}{2\pi} \, \frac{d\theta_2}{2\pi} \, .
\eeq

There are two equivalent minima, corresponding to $\{x = 1, \theta_1 = \theta_2 = 0\}$, and $\{x = 1, \theta_1 = \theta_2 = \pi\}$.  Choosing to expand around the first (and multiply the result by a factor of 2), we can define $x = 1 + \varepsilon$, and $\theta_i = \psi_i$, where $\{\varepsilon, \psi_1, \psi_2\}$ are small.  Then we can write

\beq
\langle \{2\} \rangle_{\mathrm{HS}} \approx \frac{2}{4\pi^2}\left(\frac{8\eta}{N}\right) \, g(\sigma; \eta) \, e^{1/\tilde{\tau}} \int_{0}^{\infty} e^{-p\varepsilon/\tilde{\tau}} \, d\varepsilon \, \int_{-\infty}^{\infty}\int_{-\infty}^{\infty} e^{-\frac{1}{2\tilde{\tau}}(\psi_1^2 + \psi_2^2 + \psi_1\psi_2)} \, d\psi_1 \, d\psi_2 \approx \frac{\eta\gamma(\eta)\lambda}{N} \, ,
\eeq
where $\gamma(\eta) = g(\sigma; \eta)$ is the height of the first peak in the radial distribution function for hard spheres, and 
\beq
\lambda = \frac{16\tilde{\tau}^2 \, e^{1/\tilde{\tau}}}{\pi p \sqrt{3}} \, .
\eeq
We will define our double limit formally as $p \rightarrow \infty$, $\tilde{\tau} \rightarrow 0$, with $\lambda$ fixed.  

\subsubsection{The $m$-Particle Chain}

\begin{figure}
\begin{center}
\resizebox{3in}{!}{\includegraphics{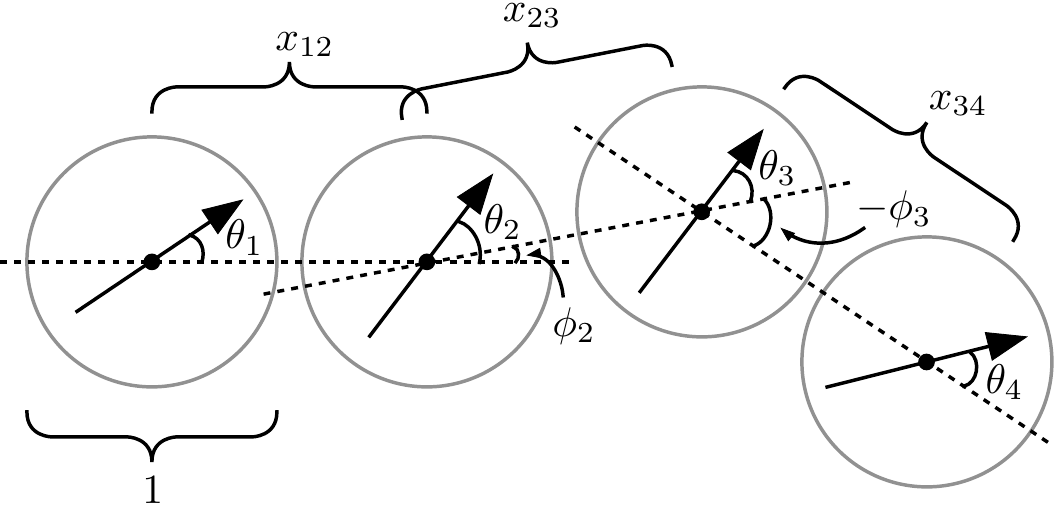}}
\caption{\label{fourchain} A diagram labeling the variables used to describe an arrangement of 4 particles in a single chain.}
\end{center}
\end{figure}

We can then extend this calculation to that for a chain diagram with $m$ particles in it.  The integral over parameter space is best expressed in terms of the variables $\{x_{i,i+1}, \theta_i, \phi_i\}$, which we show for the case $m = 4$ in figure \ref{fourchain}.

This can originally be written as
\beq
\label{eqn:mchain}
\langle \{m\} \rangle_{\mathrm{HS}} = \left(\frac{8\eta}{N}\right)^{m-1} \left[\prod_{i = 1}^{m-1}\int_{1}^{\infty} dx_{i,i+1}\right] \left[\prod_{i = 2}^{m-1} \int \frac{d\phi_i}{2\pi}\right]\left[\prod_{i = 1}^{m} \int_{0}^{2\pi} \frac{d\theta_i}{2\pi}\right] g(\sigma\vec{x}_i; \eta)
\eeq
$$
\times \left(\mathrm{Exp}\left[\frac{2\cos\theta_1\cos\theta_2 - \sin\theta_1\sin\theta_2}{2\tilde{\tau}x_{12}^{p}}\right] - 1\right) \prod_{i = 2}^{m-1} \left(\mathrm{Exp}\left[\frac{2\cos(\theta_i - \phi_i)\cos\theta_{i+1} - \sin(\theta_i - \phi_i)\sin\theta_{i+1}}{2\tilde{\tau}x_{i,i+1}^{p}}\right] - 1\right)
$$

Most significant here is the region of integration for each angle $\phi_i$, which is not explicitly written above and includes dependence on the $x_i$ variables.  However, the physical arrangements which dominate this integral will have $x_{i,i+1} \approx 1$, and in this case we get $\phi_i \in [-2\pi/3, 2\pi/3]$: the restriction to this region follows from the hard-sphere repulsion between next-nearest-neighbor particles.  With this restriction, there are only two equivalent minima, corresponding to $\{x_{i,i+1} =1, \phi_i = 0, \theta_i = 0\}$, and $\{x_{i,i+1} = 1, \phi_i = 0, \theta_i = \pi\}$.  

Expanding around the first (and multiplying the result by 2), we can define
\beq
\begin{array}{rclcl}
\psi_i & = & \theta_i, & \hspace{.5in} & i = 1, \dots, m \\
\tilde{\psi}_{i} & = & \phi_{i} - \theta_{i}, & \hspace{.5in} & i = 2, \dots, m-2 \\
\varepsilon_i & = & x_{i,i+1} -1, & \hspace{.5in} & i = 1, \dots, m-1 \, , \end{array}
\eeq
and then we obtain
\begin{samepage}
\beq
\langle \{m\}\rangle_{\mathrm{HS}} \approx 2\gamma_{m}(\eta)\left(\frac{8\eta}{N}\right)^{m-1} \, e^{(m-1)/\tilde{\tau}} \, \left[\prod_{i = 1}^{m-1} \int_{0}^{\infty} e^{-p\varepsilon_i/\tilde{\tau}}\right]\left[\prod_{i = 2}^{m-1} \int_{-\infty}^{\infty} \frac{d\tilde{\psi}_i}{2\pi}\right]\left[\prod_{i = 1}^{m} \int_{-\infty}^{\infty} \frac{d\psi_i}{2\pi}\right]
\eeq
$$
\times e^{-\frac{1}{2\tilde{\tau}}(\psi_1^2 + \psi_2^2 + \psi_1\psi_2)} \prod_{i = 2}^{m-1} e^{-\frac{1}{2\tilde{\tau}}(\tilde{\psi}_i^2 + \psi_{i+1}^2 + 2\tilde{\psi}_i\psi_{i+1})}
$$
\end{samepage}
where $\gamma_m(\eta)$ is the hard sphere distribution function for $m$ particles evaluated on the chain geometry.  This is then simply $m-1$ factors of the same integral structure which appeared in the 2-particle chain, giving us
\beq
\langle \{m\} \rangle_{\mathrm{HS}} = 2 \gamma_{m}(\eta) \left[\left(\frac{8\eta}{N}\right)\frac{e^{1/\tilde{\tau}}}{4\pi^2}\int_{0}^{\infty} e^{-p\varepsilon/\tilde{\tau}} \, d\varepsilon \, \int_{-\infty}^{\infty}\int_{-\infty}^{\infty} e^{-\frac{1}{2\tilde{\tau}}(\psi_1^2 + \psi_2^2 + \psi_1\psi_2)} \, d\psi_1 \, d\psi_2\right] ^{m-1} = 2\gamma_{m}(\eta)\left(\frac{\eta\lambda}{2N}\right)^{m-1} \, ,
\eeq

In a standard cluster expansion performed using the ideal gas as a reference system, chain diagrams factorize so that $\langle \{m\}\rangle = \langle \{2\}\rangle^{m-1}$; the fact that we are expanding instead around the hard sphere system means that the diagrams no longer behave this way.  In part this is because of the hard sphere distribution functions $\gamma_m(\eta)$.  However, even in the low density limit, where these factors become trivial, we have $\langle \{m\} \rangle_{\mathrm{HS}} \approx 2^{2-m}\langle \{2\}\rangle_{\mathrm{HS}}^{m-1}$.  The difference occurs because when considering a chain diagram $\langle \{m\}\rangle$, we neglect both the anisotropic interaction and the hard sphere interaction between particles not next to each other on the chain, which in equation \ref{eqn:mchain} would mean allowing $\phi_i \in [0, 2\pi]$.  This gives additional, non-physical minima corresponding to the chain bent back on itself at some particles.  In a full diagram expansion these would be canceled by other terms, but their presence makes it clear that we can only restrict ourselves to chain diagrams if we expand around the hard sphere system.

\subsubsection{Separated Chain Diagrams}

\begin{figure}
\begin{center}
\resizebox{4in}{!}{\includegraphics{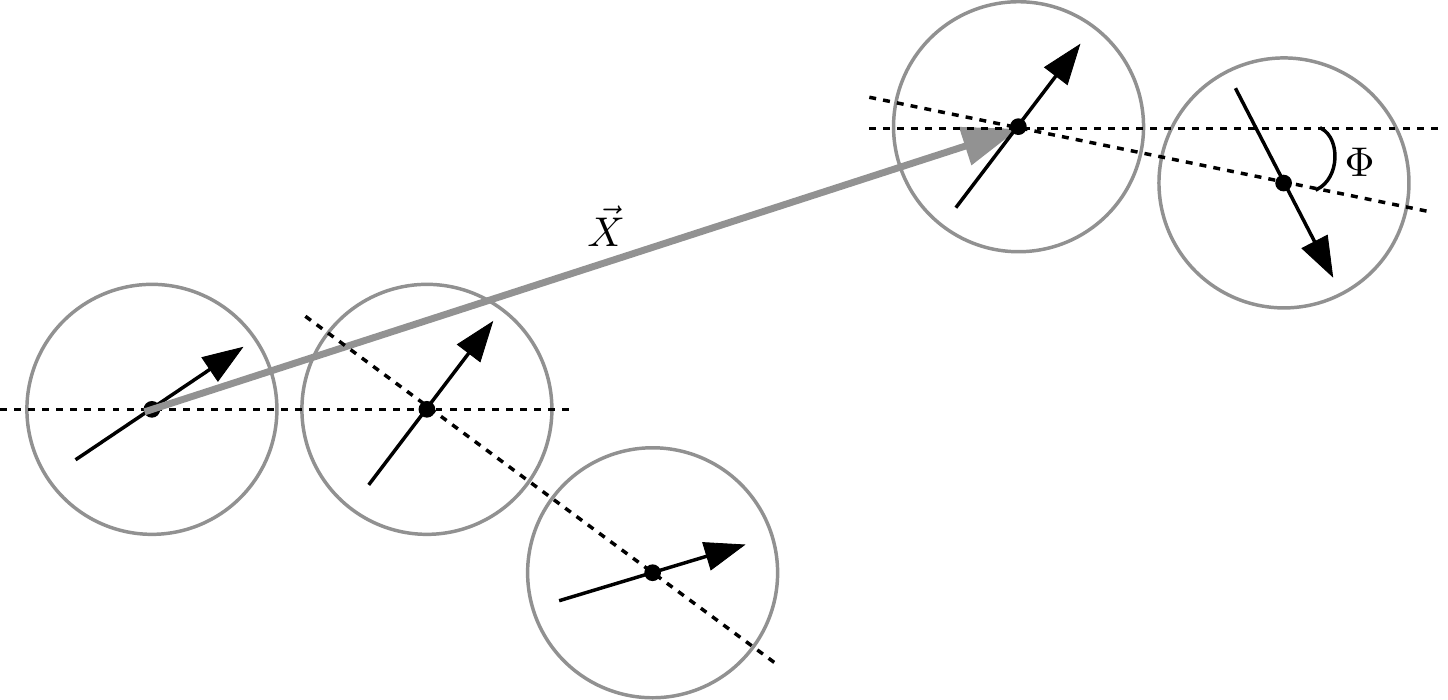}}
\caption{\label{sepchain} A diagram showing two separated chains, and the variables $\vec{X}$ and $\Phi$ giving the relative positions and orientations of the two chains.}
\end{center}
\end{figure}

Finally, consider a diagram $\langle \{m_1, \dots, m_s\}\rangle_{\mathrm{HS}}$.  This integral includes a large moduli space, corresponding to the relative positions and orientations of each sub-chain, which we can label as $\vec{X}_i$ and $\Phi_i$ for $i = 2, 3, \dots, s$.  (Figure \ref{sepchain} shows these variables for the case of two separated chains.)

However, the only part of the integrand which will depend on these moduli is the hard sphere distribution function, and apart from that the calculation factorizes into separate pieces for each sub-chain.  Thus, we obtain
\beq
\langle \{m_1, \dots, m_s\}\rangle_{\mathrm{HS}} = \prod_{i = 1}^{s} \left[2\left(\frac{\eta\lambda}{2N}\right)^{m_i-1}\right] \times \left[\prod_{i = 2}^{s} \left(\frac{8\eta}{N}\right) \int \frac{d^2\vec{X}_i \, d\Phi_i}{(2\pi)^2} \right] g(\sigma\vec{X}_i, \Phi_i; \eta) \, .
\eeq
Next we define
\beq
\gamma_{\{m_i\}}(\eta) = \left[\prod_{i = 2}^{s} \left(\frac{8\eta}{N}\right) \int \frac{d^2\vec{X}_i \, d\Phi_i}{(2\pi)^2} \right] g(\sigma\vec{X}_i, \Phi_i; \eta) \, ,
\eeq
noting that in the low density limit we would have
\beq
\gamma_{\{m_i\}}(\eta) \approx \prod_{i = 2}^{s} \left(\frac{8\eta}{N}\right) \int \frac{d^2\vec{X}_i \, d\Phi_i}{(2\pi)^2} \approx \left(\frac{8\eta}{N}\right)^{s-1} \left(\frac{A}{2\pi\sigma^2}\right)^{s-1} \approx 1 \, .
\eeq
This gives us finally
\beq
\langle \{m_1, \dots, m_s\}\rangle_{\mathrm{HS}} = 2^{s}\gamma_{\{m_i\}}(\eta) \, \left(\frac{\eta\lambda}{2N}\right)^{m-s} \, ,
\eeq
where $m = \sum_{i = 1}^{s} m_i$ is the total number of particles in the diagram. 

\subsubsection{The Free Energy at Low Density}

If we work in the low density limit, where we have $\gamma_{\{m_i\}}(0) = \gamma_m(0) = \gamma(0) = 1$, we can use these factorization rules and proceed by direct computation of corrections to the free energy.  A sum of the diagrams surviving our limit then gives
\beq
\langle e^{-U/\tau}\rangle_{\mathrm{HS}} = \sum_{n = 0}^{N} \frac{N!(N-1)!}{n!(N-n)!(N-n-1)!} \, \left(\frac{x}{2N}\right)^{n} \, .
\eeq
where $x = N\langle \{2\} \rangle_{\mathrm{HS}} = \eta\lambda$ is defined for later convenience.  The details of the combinatorial argument leading to this result appear in appendix \ref{combinatorics}.  In order to take the thermodynamic limit $N \rightarrow \infty$, we approximate the sum as an integral over $z = \frac{n}{N}$ and apply Stirling's approximation.  This leads to the expression

\beq
\langle e^{-U/\tau}\rangle_{\mathrm{HS}} = \frac{\sqrt{N}}{\sqrt{2\pi}}\int_{0}^{1} \mathrm{Exp}\left[-N\left(\ln (1 - z)^2 - z\ln\frac{x(1-z)^2}{2z} + z\right)\right] \, \frac{dz}{\sqrt{z}} \, \equiv \frac{\sqrt{N}}{\sqrt{2\pi}}\int_{0}^{1} e^{-Nf(z; x)} \, \frac{dz}{\sqrt{z}}
\eeq
(we are defining the function $f(z; x)$ in the above expression.)  Using the fact that $N$ is large, we can evaluate this expression using another saddle point approximation.  If we expand the function $f(z;x)$ near its minimum, which is located at
\beq
z_0 = \frac{1 + x - \sqrt{1 + 2x}}{x} \, ,
\eeq
 we obtain
\beq
\langle e^{-U/\tau}\rangle_{\mathrm{HS}} = \sqrt{\frac{N}{2\pi z_0}} \, e^{-Nf(z_0;x)} \, \int_{-\infty}^{\infty} e^{-\frac{Nf''(z_0;x)(z - z_0)^2}{2}} \, dz = \frac{e^{-Nf(z_0;x)}}{\sqrt{z_0f''(z_0;x)}} \, .
\eeq

When we insert this expression into the free energy consistently with the thermodynamic limit, the denominator becomes irrelevant and we obtain
\beq
\label{eqn:F}
F(\eta, \tau) = F_{\mathrm{HS}}(\eta, \tau) - \tau\ln\left\langle e^{-U/\tau}\right\rangle_{\mathrm{HS}} = F_{\mathrm{HS}}(\eta, \tau) - N\tau \Delta(\eta\lambda) \, ,
\eeq
with $\Delta(x) = -f(z_0;x)$ given by
\beq
\label{eqn:delta}
\Delta(x) = -1 - \frac{1}{x} + \frac{\sqrt{1 + 2x}}{x} + 2\ln \left[\frac{1}{2}\Big(1 + \sqrt{1 + 2x}\Big)\right] \, .
\eeq

\section{Comparisons with DHS Models}
\label{comparisons}

It is particularly interesting to consider this construction in light of the chain and polymer models used for the DHS system.  In \cite{Sear}, Sear makes the seemingly straightforward assumption that because the DHS system is known to be dominated by chain arrangements for a wide range of densities and temperatures, it is sensible to perform a cluster expansion and include only chain diagrams to construct a model valid in these regimes.  However, he works with an expansion around the ideal gas system, and uses the standard factorization rules that apply there to construct the grand partition function.  The analysis of the previous section makes it clear that this is problematic: at high temperatures (and low densities) the difference between chain diagrams $\langle \{m\}\rangle$ and $\langle \{m\}\rangle_{\mathrm{HS}}$ should be irrelevant because the excluded regions of integration for $\langle \{m\}\rangle_{\mathrm{HS}}$ are small compared to the total volume of the system, but at low temperatures this distinction becomes important because the excluded regions include additional extrema for the integrals, which do not correspond to sensible chain arrangements.

On the other hand, in \cite{OsipovTeixeiraTelodaGama, TavaresTelodaGamaOsipov} a polymer model (the ``OTT'' model) is used for the DHS system, in which the partition function for a chain of length $m$ is computed, and the free energy for ther full system is written as a sum over chains with a distribution function of chain lengths.  This distribution function is then found by minimizing the free energy.  When these techniques are applied to the system analyzed here in our double limit, we obtain the same result as equations \ref{eqn:F} and \ref{eqn:delta}.  Because the factorization rules used in our construction follow from the low temperature limit taken, we can interpret the OTT polymer model as being more correct in the low temperature limit, while the Sear chain model is more correct at higher temperatures.  The temperatures used in numerical simulations that these models are compared to typically have $\tilde{\tau} \lesssim 0.1$, where the OTT model should be more appropriate.

This model was later modified in \cite{TavaresWeisTelodaGama1999} through the inclusion of a factor of $\frac{1}{m!}$ in the partition function for a chain of length $m$, intended to account for the indistinuishability of individual monomers.  Such a factor is not standard in polymer constructions, but appears to bring the model into closer agreement with numerical results.  However, our construction shows that the original OTT model follows from a direct diagrammatic calculation, while the modifed model does not.  This suggests that the original is actually a better model for non-interacting chains, and that deviations between this model and the numerical results for the DHS system are more likely a result of chain-chain interactions.

\section{Finite Density Effects}
\label{finitedensity}

The arguments which lead to the conclusion that in our double limit the system is dominated by chains do not require we work at low density.  However, in order to use this result to compute the free energy, we assumed a low enough density that we could drop the factors of the hard sphere distribution function: $\gamma_{\{m_i\}}(\eta) \approx 1$.  On the other hand, it is possible to include some finite density effects by assuming that the hard sphere distribution functions approximately factor according to the rule $\gamma_{\{m_i\}}(\eta) \approx \gamma(\eta)^{m}$.  This amounts to the assumption that the dominant finite density effect in the probability of finding hard spheres in chain arrangements comes from the separate probability of finding each adjacent pair of particles next to each other.\footnote{This is not quite the same as a superposition approximation for the distribution functions, which would also include factors from pairs of particles not next to each other.}  Although we do not expect the results of making this approximation to be in perfect quantitative agreement with the exact results, it should provide insight into the way finite density hard-sphere effects influence the results.  

\begin{figure}
\begin{center}
\resizebox{3in}{!}{\includegraphics{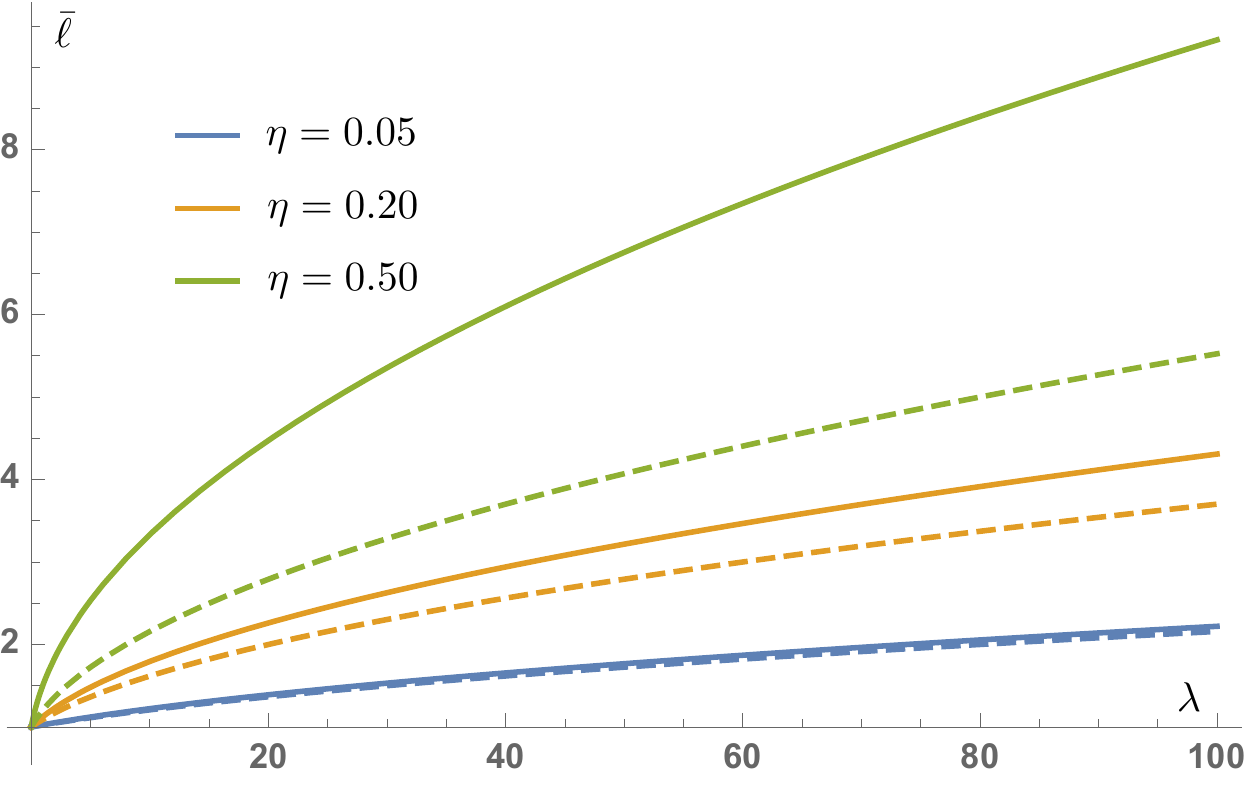}}
\caption{\label{lengthplot} A graph of typical chain length $\ell$ vs. $\lambda$ for values of $\eta = 0.05$, $\eta = 0.20$, and $\eta = 0.50$.  The dashed lines give the same results, but with $\gamma(\eta) \rightarrow 1$.}
\end{center}
\end{figure}

Conveniently, these effects are now being taken into account only through the inclusion of the height of the first peak in the radial distribution function for hard spheres in 2 dimensions, $\gamma(\eta)$, which has been thoroughly studied.  In what follows, we will use the approximation from \cite{Luding},
\beq
\gamma(\eta) \approx \frac{1 - \frac{7\eta}{16}}{(1 - \eta)^2} - \frac{\eta^3}{128(1 - \eta)^4} \, ,
\eeq
which is known to be in excellent agreement with numeric results up to around $\eta \lesssim 0.5$.  Then, we modify equation \ref{eqn:F} by using $x = \eta\gamma(\eta)\lambda$, so that we have
\beq
F(\eta, \tau) \approx F_{\mathrm{HS}}(\eta, \tau) - N\tau \Delta\Big(\eta\gamma(\eta)\lambda\Big) \, .
\eeq
We can calculate the average energy per particle in the system using $\langle \tilde{u} \rangle = -\frac{\tau^2}{Nu_0}\frac{\partial (F/\tau)}{\partial \tau}$, which gives
\beq
\langle \tilde{u} \rangle = -\eta \gamma(\eta) \lambda \, \Delta'\Big(\eta\gamma(\eta)\lambda\Big) = -\frac{\eta \gamma(\eta)\lambda}{1 + \eta \gamma(\eta)\lambda + \sqrt{1 + 2\eta \gamma(\eta)\lambda}} \, .
\eeq
For a straight chain of $\ell$ particles, the energy per particle in the chain will be $-\frac{(\ell - 1)}{\ell}$, so we can define the typical chain length in an arrangement as $\bar{\ell} = \frac{1}{1 + \langle \tilde{u}\rangle}$, which gives
\beq
\label{eqn:length}
\bar{\ell} = \frac{1 + \sqrt{1 + 2\eta\gamma(\eta)\lambda}}{2} \, .
\eeq  
Figure \ref{lengthplot} shows this typical length as a function of $\lambda$, for different values of $\eta$.  Unsurprisingly, chains are longer for larger values of $\lambda$ or $\eta$, and approach infinite length as $\lambda \rightarrow \infty$.  In order to better examine the influence of finite density hard sphere effects, we also show the results with the replacement $\gamma(\eta) \rightarrow 1$ (as dashed lines).

\begin{figure}
\begin{center}
\resizebox{3in}{!}{\includegraphics{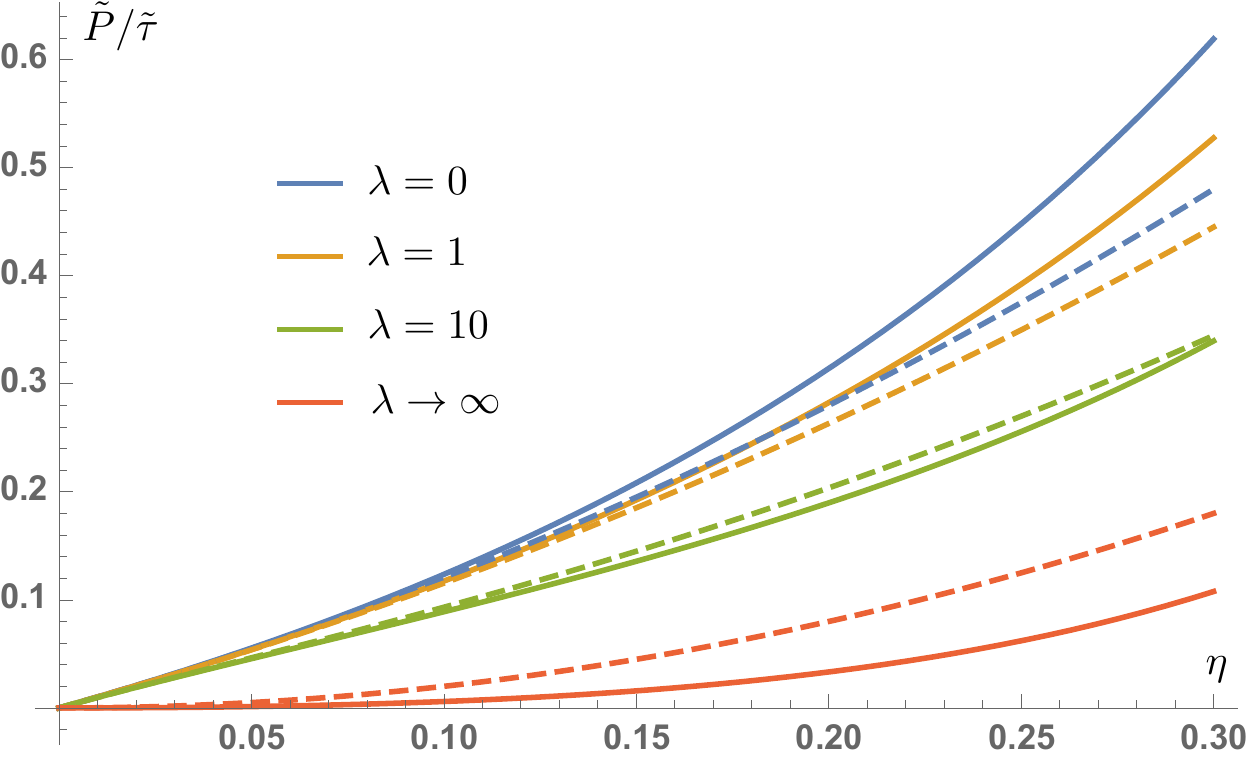}} 
\caption{\label{pressureplot} A graph of $\tilde{P}/\tilde{\tau}$ vs. $\eta$ for values of $\lambda = 0$, $\lambda = 1$, $\lambda = 10$, and $\lambda \rightarrow \infty$.}
\end{center}
\end{figure}

Next we consider the equation of state for this system.  Remembering that for a pure system of hard spheres we have $P_{\mathrm{HS}} = \frac{N\tau}{A}\left[1 + 2\eta \gamma(\eta)\right]$, we can use $P = -\frac{\partial F}{\partial A}$ to write
\beq
\label{eqn:P}
\tilde{P} = \left(\frac{\pi\sigma^2u_0}{4}\right)P = \eta\tilde{\tau}\left[1 + 2\eta \gamma(\eta) - \frac{\eta \gamma(\eta)\lambda\left(1 + \frac{\gamma'(\eta)}{\gamma(\eta)}\right)}{1 + \eta \gamma(\eta)\lambda+ \sqrt{1 + 2\eta \gamma(\eta)\lambda}}\right] \, .
\eeq
Figure \ref{pressureplot} shows a graph of $\tilde{P}/\tilde{\tau}$ vs. $\eta$ for various values of $\lambda$, again with the same results with $\gamma(\eta) \rightarrow 1$ in dashed lines.  Unsurprisingly, no liquid-gas phase transition occurs in this system (the double limit taken completely suppresses the Tlusty-Safran mechanism of \cite{TlustySafran}).  Furthermore, if we consider the low density limit of this equation of state (where hard sphere effects disappear), we recover the expression for an ideal gas of chains: $P = \frac{n\tau}{A}$, where $n = \frac{N}{\bar{\ell}}$ is the number of chains present.  More interestingly, we see that although for most values of the parameters finite density effects increase the pressure (as they would for the pure hard sphere system), for large values of $\lambda$ and moderate values of $\eta$ they can decrease it due to their influence on chain formation.

\section{Conclusions and Future Directions}
\label{conclusions}

In this work we have analyzed a 2-dimensional system of hard spheres subject to a short-ranged anisotropic interaction at low temperatures.  By performing a cluster expansion around a hard sphere system and using a saddle-point approximation for each term in the expansion, we have argued that a double limit exists where the temperature approaches zero as the interaction becomes increasingly short-ranged, and that in this limit the only structures which contribute to the free energy of the system are chains.  This is also connected to the fact that for any interaction range parameter $p > 3$, the interaction is sufficiently short-ranged that the interaction potential will not contribute to the free energy beyond the correlation length.  

We then showed that the contributions from these structures were contained in chain diagrams, analyzed the factorization properties of these diagrams, and used this to find a closed form expression for the free energy of the system at low densities.  We compared our construction to those used in three different models for the chain-dominated regime of the DHS system, and concluded that we could reproduce the results of the OTT polymer model, suggesting that it is more correct at low temperatures while the Sear model would be more correct at higher temperatures.  Finally, by making simplifying assumptions regarding the factorization rules for hard sphere distribution functions, we were able to incorporate some finite density effects into the model, and analyze their effects on chain length and the equation of state.

In the future, it would be interesting to analyze chain diagrams $\langle \{m\}\rangle_{\mathrm{HS}}$ at finite temperatures, determine if approximate factorization rules might be found, and attempt to incorporate them into the free energy.  This might produce a way of interpolating between the OTT polymer model and the Sear chain model, possibly producing better agreement with finite temperature numerical data.  Furthermore, the simulation data these models were compared to invariably used a physically realistic dipole interaction, with $p = 3$ (even in later works when the particles were confined to a plane).  It would be useful to have simulation data for different values of $p$, to compare the various models to.  In particular, we would expect that a series of simulations run for different values of $p$ and $\tilde{\tau}$ but the same value of $\lambda$ would reveal that as $p$ is increased the dominance of chain structures is exaggerated.  It might then be possible to confirm which model is more appropriate for a system of non-interacting chains, and how much of the deviations between existing models and data is due to chain-chain interactions.  It would also be interesting to explore a possible connection between the ideas explored in this work relating polymer theory to a diagrammatic expansion, and work in \cite{HoyeStellLee} relating polymer theory to the random-walk structures in the Ornstein-Zernike equations.

Finally, it would be interesting to apply the arguments in section \ref{generalities} to a dipole system in an arbitrary number of dimensions.  We might expect that the simplifications associated with large values of $p$ would carry over to a self-consistent theory of hard sphere dipoles in a large number $D$ dimensions, where we would naturally have 
\beq
u_{12} = \frac{u_0\sigma^{D}}{(D-1)r_{12}^{D}}\Big[\hat{m}_1\cdot\hat{m_2} - D(\hat{m}_1\cdot\hat{r}_{12})(\hat{m}_2\cdot\hat{r}_{12})\Big] \, .
\eeq  
In this case, the modification to the structure of the anisotropic interaction would also favor chain formation as $D$ becomes large: the interaction energy associated with a head-to-tail arranement of particles would be $u_{12} = -u_0$, while that of the anti-parallel side-by-side arrangement would be $u_{12} = -\frac{u_0}{D-1}$.  Many physical systems are known to simplify dramatically in the ``infinite dimensional'' limit, and this work suggests the DHS system would as well.

\acknowledgements{The author would like to thank the UAlbany theory group for interesting discussions, as well as Heather Watson and Jef Wagner, for serving as a continual sounding board.}

\appendix

\section{Summation of Diagrams}
\label{combinatorics}

We are interested in performing the sum
\beq
\label{eqn:sum}
\langle e^{-U/\tau} \rangle_{\mathrm{HS}} = 1 + \frac{N!}{2(N-2)!}\langle f_{12}\rangle_{\mathrm{HS}} + \frac{N!}{2(N-3)!}\langle f_{12}f_{23}\rangle_{\mathrm{HS}}  + \frac{N!}{8(N-4)!}\langle f_{12} f_{34} \rangle_{\mathrm{HS}} + \cdots
\eeq
where the only terms included in the sum are chain diagrams and diagrams consisting of separated chains, and where we assume the factorization rules
\beq
\langle f_{12}f_{23}\cdots f_{n,n+1}\rangle_{\mathrm{HS}} = \frac{1}{2^{n-1}}\langle f_{12}\rangle_{\mathrm{HS}}^{n}
\eeq
\beq
\langle f_{12}f_{23}\cdots f_{n,n+1}f_{n+2,n+3}\cdots f_{m,m+1}\rangle_{\mathrm{HS}} = \langle f_{12}f_{23}\cdots f_{n,n+1}\rangle_{\mathrm{HS}}\langle f_{n+2,n+3}\cdots f_{m,m+1}\rangle_{\mathrm{HS}}
\eeq
(that is, separated chains factor straightforwardly, but Meyer functions inside chains generate additional terms of $\frac{1}{2}$ when they are factorized.)

Using these rules, we should be able to collect terms based on the number $n$ of Meyer functions they include, and rewrite the sum in equation \ref{eqn:sum} as
\beq
\langle e^{-U/\tau}\rangle_{\mathrm{HS}} = \sum_{n = 0}^{N} C(N, n)\langle f_{12} \rangle_{\mathrm{HS}}^{n}
\eeq
where the coefficients $C(N, n)$ are determined both by counting diagrams and by including factors of $\frac{1}{2}$ coming from chain factorization.  In order to find these coefficients, consider the following:

\begin{enumerate}

\item We clearly have $C(N, 0)= 1$, and $C(N, 1) = \frac{N(N-1)}{2}$, which is simply the number of choices for a single Mayer function in a system of $N$ particles.  It is convenient to think of this as the number of ways we can take a set of $N$ points, and connect exactly one pair of points via a link.

\item In determining $C(N, 2)$, suppose that we already have one link between two points (label them $1$ and $2$ without loss of generality), and we must form another.  The new link may be entirely separate from $1$ and $2$, or it may include either $1$ or $2$, but not both.  Suppose it includes $1$, and a new point we label $3$.  The diagram created this way includes a chain of length $3$, and when we use the factorization rules given above to rewrite it as a product of two chains of length $2$, we introduce a factor of $\frac{1}{2}$.  Similarly, the new link could inlcude $2$ and the new point $3$, and this would again introduce a factor of $\frac{1}{2}$.  If we combine these two options, the contribution to $C(N, 2)$ is the same as if we treat the existing points $1$ and $2$ as if they have been fused into a single point $1'$.  Thus, the total contribution to $C(N, 2)$ from forming the second link is the same as the number of ways we could form a link between two points in the set $\{1', 3, 4, \dots, N\}$ (which has $N-1$ members.)  We then obtain
\beq
C(N, 2) = \frac{1}{2!}\left[\frac{N(N-1)}{2}\right]\left[\frac{(N-1)(N-2)}{2}\right]
\eeq 
where the factor of $\frac{1}{2!}$ corrects for over-counting by forming the links in a particular order.

\item This logic than then be extended to find $C(N, n)$: Having formed the first $n-1$ links, we want to account for the number of ways to put in the last link.  This new link can connect to an already existing chain, but if it does it must connect to one of the two ends (otherwise you end up with a non-chain diagram).  Furthermore, by connecting to an existing chain, you introduce an additional factor of $\frac{1}{2}$ when you factorize that chain.  Thus, the total contribution to $C(N, n)$ from these possibilities is the same as if you treat any existing chains as single points.  After having formed $n-1$ links, this means you are counting the number of ways of forming a link between $N - n + 1$ points.  Thus, we should have
\beq
C(N, n) = \frac{1}{n!}\left[\frac{N(N-1)}{2}\right]\left[\frac{(N-1)(N-2)}{2}\right]\cdots\left[\frac{(N-n+1)(N-n)}{2}\right] 
\eeq
and finally
\beq
C(N, n) = \frac{N!(N-1)!}{2^{n}n!(N-n)!(N-n-1)!} \, .
\eeq
(This formula has also been explicitly checked against the diagrammatic expansion out to $n = 10$, with the aid of Mathematica.)

\end{enumerate}

\bibliography{dipolearticles}{}
\bibliographystyle{ieeetr}

\end{document}